\documentclass{icrc2009}

\usepackage{graphicx}   
\usepackage{caption}    
\usepackage[font=footnotesize]{subfig} 
\usepackage{fixltx2e}
\usepackage{cite}
\usepackage{url}

\newcommand{\shorttitle}[1]%
{\markboth{Proceedings of the 31\MakeLowercase{$^{st}$} ICRC, {\L}\'{o}d\'{z} 2009}{#1} }
\newcommand{\etal}{\MakeLowercase{\textit{et al. }}} 

\begin{document}
\title{MAGIC observation of GRB\,090102}

\author{\IEEEauthorblockN{	M. Gaug\IEEEauthorrefmark{1},
    S. Covino\IEEEauthorrefmark{2},
    M. Garczarczyk\IEEEauthorrefmark{3},
    A. Antonelli\IEEEauthorrefmark{4},
    D. Bastieri\IEEEauthorrefmark{5},
    A. La Barbera\IEEEauthorrefmark{4},\\
    J. Becerra-Gonzalez\IEEEauthorrefmark{3},
    A. Carosi\IEEEauthorrefmark{4},
    N. Galante\IEEEauthorrefmark{6},
    F. Longo\IEEEauthorrefmark{7},
    V. Scapin\IEEEauthorrefmark{8},
    S. Spiro\IEEEauthorrefmark{4}}
  for the MAGIC collaboration \\
  
						\\

\IEEEauthorblockA{\IEEEauthorrefmark{1}Instituto de Astrof\'isica de Canarias, via L\'actea s/n, 38205 La Laguna, Tenerife, Spain}
\IEEEauthorblockA{\IEEEauthorrefmark{2}INAF / Brera Astronomical Observatory, Via Bianchi 46, 23807, Merate (LC), Italy}
\IEEEauthorblockA{\IEEEauthorrefmark{3}IFAE, Edifici Cn, Campus UAB, 08193 Bellaterra, Spain}
\IEEEauthorblockA{\IEEEauthorrefmark{4}INAF / Rome Astronomical Observatory, Via Frascati 33, 00044, Monte Porzio (Roma), Italy}
\IEEEauthorblockA{\IEEEauthorrefmark{5}Universit\`a di Padova and Istituto Nazionale di Fisica Nucleare (INFN), 35131, Padova, Italy}
\IEEEauthorblockA{\IEEEauthorrefmark{6}Max-Planck-Institut f\"ur Physik, F\"ohringer Ring 6, 80805 M\"unchen, Germany}
\IEEEauthorblockA{\IEEEauthorrefmark{7}Dipartimento Fisica and INFN Trieste, 34127 Trieste, Italy}
\IEEEauthorblockA{\IEEEauthorrefmark{8}Universit\`a di Udine, and INFN Trieste, 33100 Udine, Italy}
}

\shorttitle{Gaug \etal MAGIC observation of GRB\,090102}
\maketitle

\begin{abstract}
On January 2, the MAGIC-I Telescope observed GRB090102 (z=1.55) under particularly good observation conditions. 
Using the recently upgraded MAGIC-1 sum trigger system, upper limits down to below 50~GeV have been obtained. 
This is the first time that the new trigger system was fully exploited for a Gamma-ray Burst (GRB) observation 
and shows the capabilities of the MAGIC observatory for future follow-up observations.
\end{abstract}

 
\section{Introduction}
GRBs are known to emit radiation up to the detector sensitivity limits of GRB satellites. 
Photons at energies $>$10~GeV have been detected so far~\cite{GRB090816}, without any sign of apparent cutoffs in the energy spectrum. 
Gamma-rays of even higher energy would allow for valuable insights in the secondary emission processes: 
Inverse Compton (IC) and/or hadronic emission scenarios which lead to a clearer picture of the circumburst medium and general energetics 
involved in the prompt emission phase. 
Whereas current satellite experiments are rather limited by their effective areas when it comes to energies in the GeV regime, 
the current generation of Imaging Atmospheric Cherenkov Telescopes (IACTs), deploys full sensitivities only above typically 200~GeV. 
At these energies, $\gamma$-ray emitting sources have to lie at small redshifts to be detectable, 
given the absorption of $\gamma$-rays by the Extragalactic Background Lights (EBL). 
GRBs are located at redshifts typically around or higher than a redshift of $z\!=\!1$, making it mandatory to lower the energy threshold of IACTSs 
to energies well below 100~GeV. This goal was recently achieved by the MAGIC collaboration using an upgraded sum trigger system and demonstrated 
with the observation of the Crab pulsar at 25~GeV~\cite{Crab}. Consequently, the new trigger system is also used for most GRB observations 
carried out with the MAGIC telescope. 
Among these observations, GRB090102 stakes out by the fact that it could be observed under particularly low zenith angles allowing for the lowest possible thresholds.

\section{MAGIC observations}

GRB090102 was detected by the \textit{Swift} satellite on January 1$^\mathrm{st}$, 2009 at 02:55:45\,UT~\cite{Mag09} and also by the Konus-Wind satellite at 
02:55:36\,UT~\cite{Gol09}. The prompt emission lasted for about 30\,s, an X-ray and an optical counterpart were discovered and followed-up by several groups. 
Spectroscopy was carried out allowing to derive a redshift of $z = 1.547$~\cite{NOT09}. Using this redshift, Konus-Wind derived an isotropic energy release 
of $E_{iso} \sim 1.9 \cdot 10^{53}\mathrm{erg}$~\cite{Gol09}. The bright multi-wavelength emission and good observational conditions for the Swift instruments 
made that the Swift collaboration declared GRB090102 as ``burst of interest''\,\footnote{Some GRBs with especially good multi-wavelength coverage at high energies are declared ``burst of interest'' by the Swift to encourage further multi-wavelength observations.}. 
 
\begin{figure}[!t]
\centering
\includegraphics[width=2.5in]{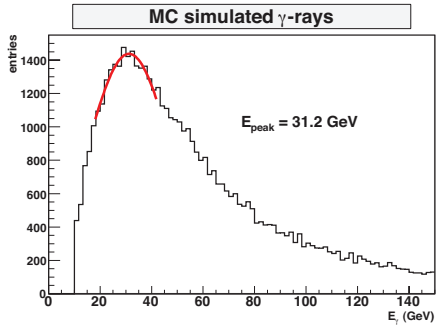}
\caption{The energy threshold for this analysis of GRB090102.}
\label{fig1}
\end{figure}

The MAGIC-I telescope started observation at 03:14:52~UT, 1161~s after the onset of the burst, but well after the prompt emission phase. 
Observations were carried out under excellent observation conditions and at a very low zenith angle, ensuring a low energy threshold. 
The observation with MAGIC-I started at a zenith angle of Z$_{\rm d} = 5^\circ$, reaching Z$_{\rm d} = 52^\circ$ after 13149~s of observation. 
One night later, extensive OFF data were collected under the same observational conditions. 
OFF data are taken with the telescope pointing close to the original source location, without the source having in its field-of-view and are used
 to estimate the background from hadronic cosmic rays triggers.
No significant excess of $\gamma$-rays above background was detected. Upper limits above 80~GeV have already been reported elsewhere~\cite{MAGICGCN}, 
while we concentrate here on the analysis of the very low-energy part of the data. As the trigger energy threshold depends very sensitively on the telescope zenith angle, 
only data with Z$_{\rm d} < 25^\circ$ were used here, ensuring an analysis threshold around 30~GeV (see figure~\ref{fig1}). These data cover 5919~s of observation time. 
Differential upper limits with 95\% CL for six energies were derived as shown in table~\ref{uls}. 
These limits include a 30\% systematic uncertainty on the overall telescope efficiency.

\section{Results}

XRT data show at the time of the MAGIC observation an average (unabsorbed) flux of $6.6 \cdot 10^{-11}\mathrm{erg}\,\mathrm{cm^{-2}}\,\mathrm{s^{-1}}$ 
in the 0.3 to 10~keV energy regime, the time averaged spectrum yields a spectral photon index of $1.87 \pm 0.16$~\cite{XRT}. 
A simple extrapolation to the energies accessible by MAGIC, made only for illustrative reasons, gives
fluxes ranging from $4 \cdot 10^{-11}\ \mathrm{to}\ 10^{-8}\,\mathrm{erg}\,\mathrm{cm^{-2}}\,\mathrm{s^{-1}}$ for our lowest energy bin from 25 to 50~GeV. 
Under the assumptions of a similar burst environment ($n \sim 1\,\mathrm{cm^{-3}}$, $\epsilon_e \sim 0.1$, $\epsilon_B\sim 0.01$), an SSC scenario, 
like the one shown in~\cite{GRB080430}, predicts a flux of about $10^{-11}\,\mathrm{erg}\,\mathrm{cm^{-2}}\,\mathrm{s^{-1}}$ at 45~GeV and $T_0+1.2\,\mathrm{ks}$, 
which is not far from the upper limit shown in the first energy bin, seen in table~\ref{uls}. 
Note that this burst is much less studied than GRB080430, and therefore all assumptions on the circumburst medium, magnetic energy density, etc. are only rudimentary, 
and variations of these parameters may modify the result considerably.

\begin{table}[h!] 
\centering 
\vspace{0.5cm}
\begin{tabular}{|c|c|c|} 
\hline 
{$\langle E \rangle$} & 
{fluence limit} & 
{average flux limit} \\
{(GeV)} & 
{$(\mathrm{erg}\cdot\mathrm{cm}^{-2})$} &
{$(\mathrm{erg}\cdot\mathrm{cm}^{-2}\cdot\mathrm{s^{-1}})$} \\ \hline 
  43.9 &  5.2$\cdot 10^{-6}$ &  8.7$\cdot 10^{-10}$ \\
  57.3 &  8.9$\cdot 10^{-7}$ &  1.5$\cdot 10^{-10}$ \\
  90.2 &  1.8$\cdot 10^{-6}$ &  3.1$\cdot 10^{-10}$ \\
 137.2 &  1.3$\cdot 10^{-6}$ &  2.2$\cdot 10^{-10}$ \\
 209.4 &  9.5$\cdot 10^{-7}$ &  1.6$\cdot 10^{-10}$ \\
 437.6 &  1.8$\cdot 10^{-7}$ &  0.3$\cdot 10^{-10}$ \\
\hline 
\end{tabular} 
\caption{Upper limits for the emission of GRB090102, corresponding to $5919\,\mathrm{s}$ of the MAGIC-I observation from 03:14:52~UT to 04:53:32~UT.}
\label{uls}
\end{table}

This simplistic treatment does not take into account higher orders of IC scattering of X-ray photons and/or hadronic emission processes 
since some of the necessary parameters are not available for this burst. One can see a recent review of possible scenarios in~\cite{Cov09} 
and an example for the case of GRB080430, compared to MAGIC upper limits, in these proceedings~\cite{GRB080430}.

\section{Extragalactic Background Light}

Different models predict a wide range of optical depths at $z \sim 1$, ranging from about 1 up to 6~\cite{FaPi08,Fran08,Gilm09,Knei04}. 
Recently, the MAGIC collaboration published an observational result~\cite{Alb08} suggesting the EBL attenuation could be much lower than previously assumed. 
A quasar located at a redshift of $z=0.536$ could be clearly detected at energies well above 150~GeV, opening such the window to the universe observable by IACTs. 
This means that at the redshift of GRB090102, $z\! \sim \! 1.5$, and at the MAGIC-1 threshold energy, E $\sim 30$\,GeV, an optical depth, 
$\tau$, not far from unity is possible. These statements are strengthened by the discovery of a $>$10~GeV emission component by LAT detector onboard the Fermi satellite, 
from a burst with much higher redshift~\cite{GRB090816}. 
In total, even after taking into account a possible attenuation of a factor 10 to 100, our limits start to be sensitive to possible IC emission components of GRBs.

\section{Light Curves}

\begin{figure}[h!]
 \centering
\includegraphics[width=2.5in]{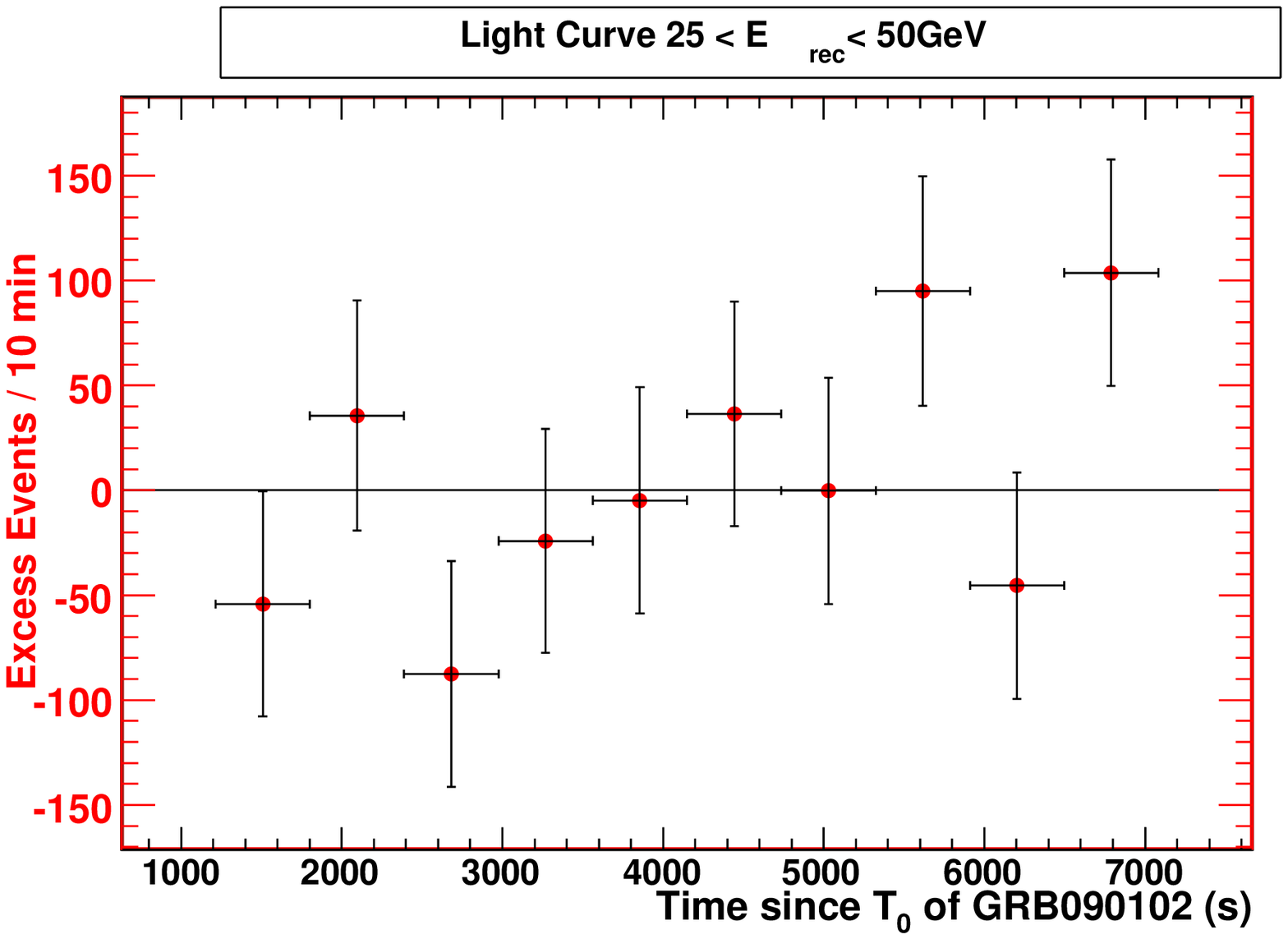} 
\includegraphics[width=2.5in]{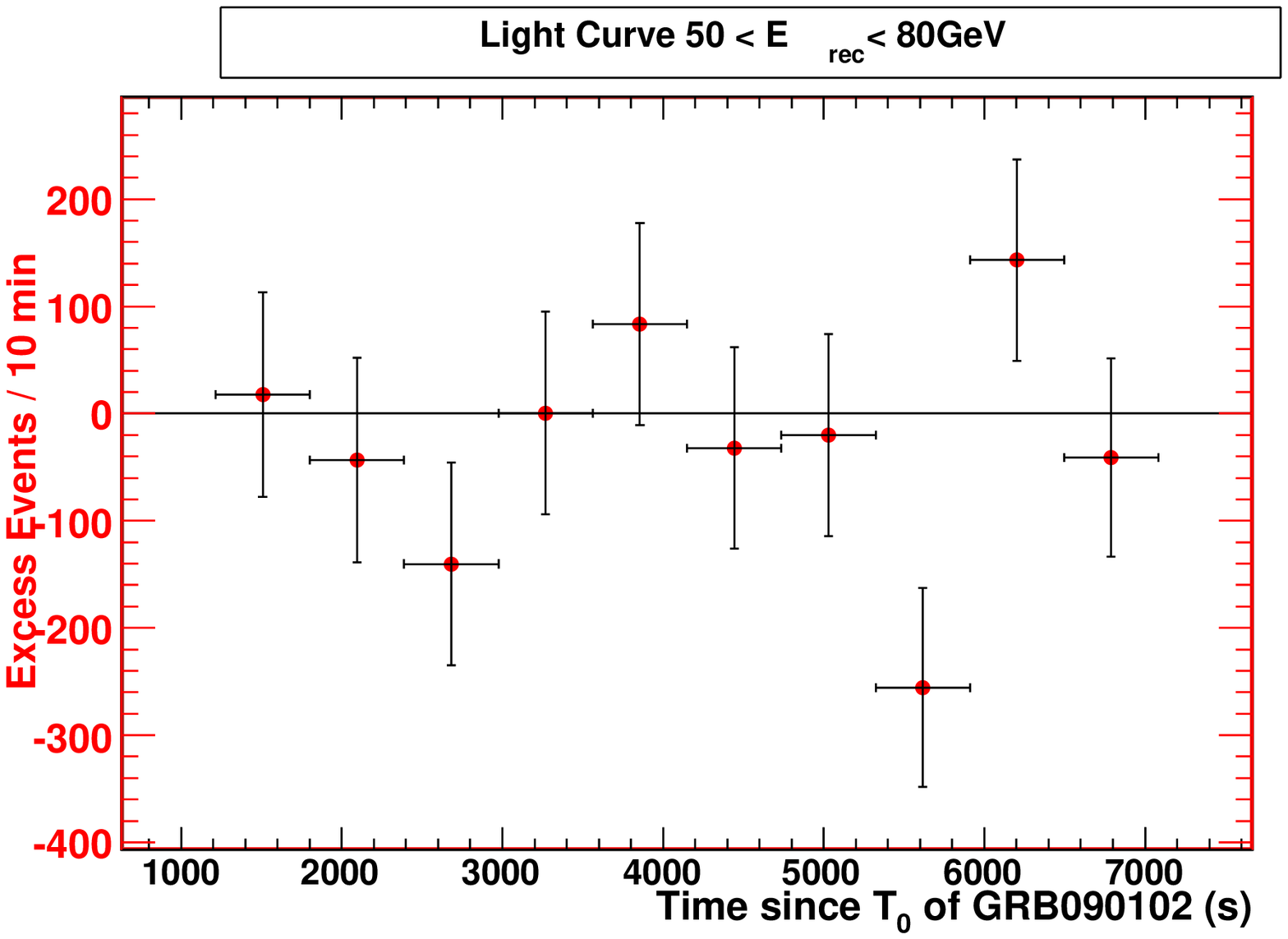} 
\includegraphics[width=2.5in]{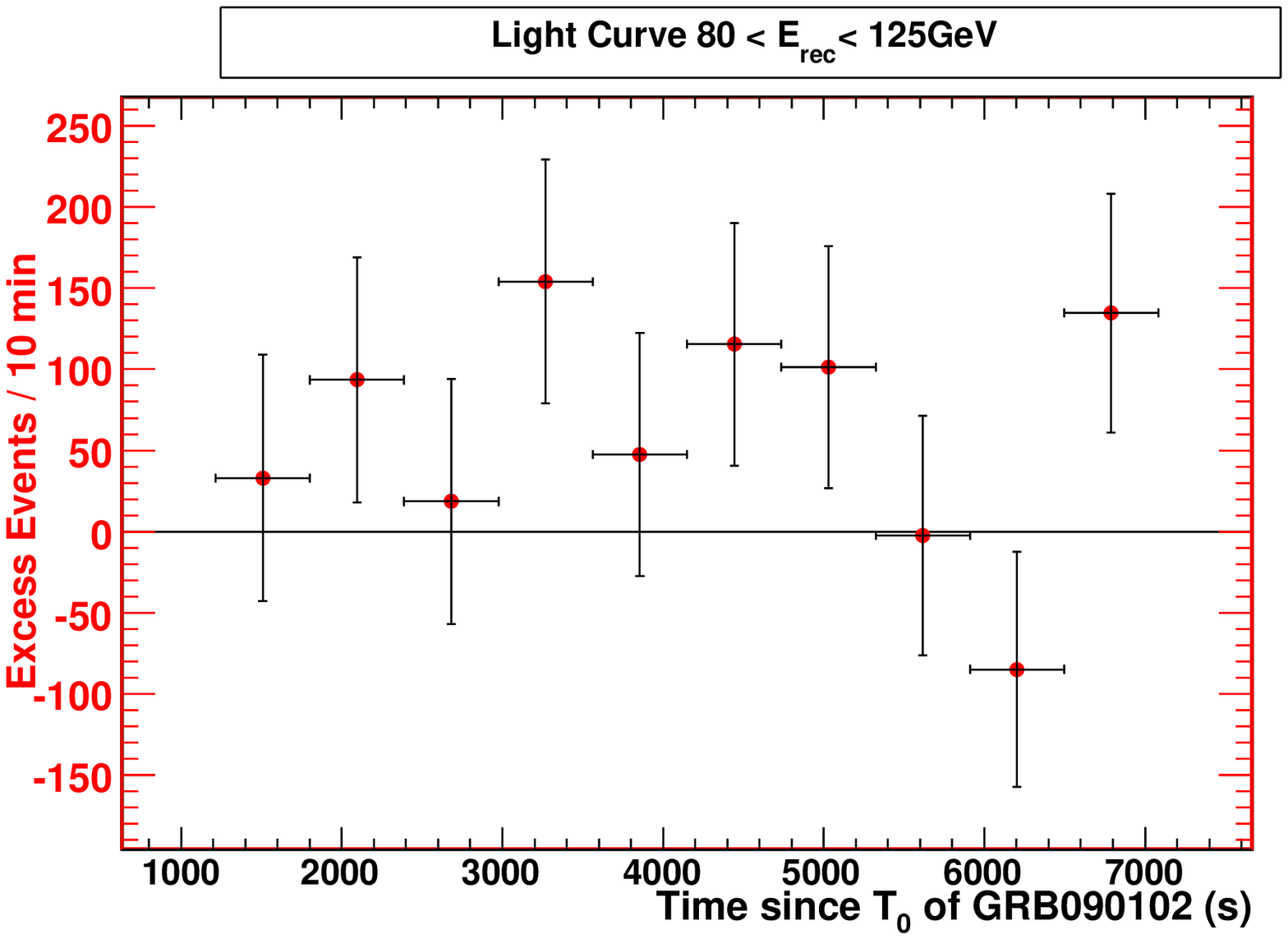} 
 \caption{The excess event rates in bins of 10 minutes, for the three lower energy bins. All bins are compatible with fluctuations of the background.}
 \label{fig2}
\end{figure}

Excess event rates vs. observation time in bins of 10 minutes are shown in figure~\ref{fig2} for the three lowest energy bins. The entries fluctuate around zero 
and are all compatible with fluctuations of the background from hadronic cosmic rays triggers.
Unfortunately, this burst did not show any X-ray flare during the early afterglow which would otherwise allow for correlation studies between the X-ray and 
VHE $\gamma$-ray emission. 

\section{Conclusions}

The MAGIC-1 Telescope has undergone several hardware upgrades~\cite{GRB} during the last year, which significantly enhance the chances to detect 
the $>$10~GeV $\gamma$-ray emission component of GRBs. GRB090102 was the first GRB afterglow observation made with the new sum trigger, 
under excellent observation conditions, which allowed to derive upper limits well below 50~GeV. Sensitivity for $\gamma$-ray emission at these low energies 
is a prerequisite to detect emission from bursts at moderate and high redshifts. Although in the case of GRB090102, our limits are still compatible with a 
general SSC model under very rude assumptions, a slightly closer and better studied burst should soon allow for the detection of a GeV counterpart, 
in the best case in a joint observation together with LAT.

\subsection*{Acknowledgements}
The collaboration thanks the Instituto de Astrof\'ısica de Canarias for the excellent working conditions at the Observatorio
del Roque de los Muchachos in La Palma, as well as the German BMBF and MPG, the Italian INFN and Spanish MCINN. 
This work was also supported by ETH Research Grant TH 34/043, by the Polish MniSzW Grant N N203 390834, and by the YIP of the Helmholtz
Gemeinschaft.

\end{document}